\RequirePackage{fix-cm}
\documentclass[smallcondensed]{svjour3}     % onecolumn (ditto)
\smartqed  % flush right qed marks, e.g. at end of proof
\usepackage{bm}
\usepackage{multirow}
\usepackage{graphicx}
\usepackage{algorithm}
\usepackage{algorithmic}
\usepackage{amsmath}

\usepackage{url}
\usepackage{subfigure}
\begin{document}

\title{Predicting the Popularity of Topics based on User Sentiment in Microblogging Websites%\thanks{Grants or other notes
}


\author{Xiang Wang \and
        Chen Wang \and
        Zhaoyun Ding \and
        Min Zhu \and
        Jiumin Huang
}


\institute{Xiang Wang, Min Zhu \at
              Academy of Ocean Science and Engineering, National University of Defense Technology, Changsha, 410073, China \\
              Tel.: +86-18975898277\\
              \email{xiangwangcn@nudt.edu.cn}           %  \\
           \and
           Zhaoyun Ding \at
              College of Information System and Management, National University of Defense Technology, Changsha, 410073, China
           \and
           Chen Wang, Jiumin Huang \at
           College of Computer, National University of Defense Technology, Changsha, 410073, China
}

\date{Received: Aug. 06, 2016 / Accepted: date}

\maketitle

\begin{abstract}
Behavioral economics show us that emotions play an important role in individual behavior and decision-making. Does this also affect collective decision making in a community? Here we investigate whether the community sentiment energy of a topic is related to the spreading popularity of the topic. To compute the community sentiment energy of a topic, we first analyze the sentiment of a user on the key phrases of the topic based on the recent tweets of the user. Then we compute the total sentiment energy of all users in the community on the topic based on the Markov Random Field (MRF) model and graph entropy model. Experiments on two communities find the linear correlation between the community sentiment energy and the real spreading popularity of topics. Based on the finding, we proposed two models to predict the popularity of topics. Experimental results show the effectiveness of the two models and the helpful of sentiment in predicting the popularity of topics. Experiments also show that community sentiment affects collective decision making of spreading a topic or not in the community.

\end{abstract}

\section{Introduction}

Predicting is a forever interesting research point in human history. With the rapid development of Web 2.0 applications, massive online users' personal information appears in these applications and makes the predicting of online content possible. We take the Chinese microblogging website SINA Weibo for instance. It is very similar to Twitter in United States and is one of the most popular sites in China. There are more than 503 million registered users and about 100 million messages posting each day on SINA Weibo~\footnote{SINA Weibo. Wikipedia. https://en.wikipedia.org/wiki/Sina\_Weibo}. Given such amazing large number of information spreading between a large number of users, microblogging website is a channel of information propagation of the topics in the real world, and it also influences the development of the related events, activities in the real world. For example, the hot event of ``Guo Meimei" in June 2011 affects the credit of the Red Cross Society of China deeply, which received much less donation after the event. Guo Meimei was authenticated as the general manager of Red Cross Commerce, showed pictures of her lavish lifestyle of driving Mercedes and owning a big mansion on SINA Weibo. Her lavish lifestyle drew skepticism from people who questioned if her wealth came from the Red Cross Society of China. So it is important to predict the popularity of topics.
\begin{figure}
  \centering
  \includegraphics[width=7.0cm]{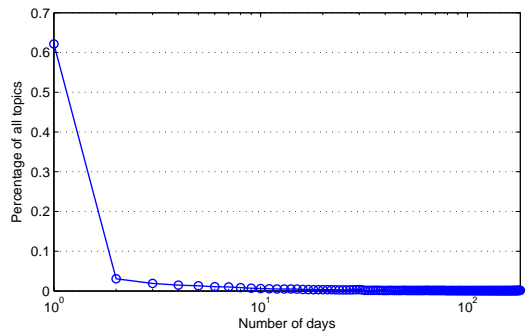}\\
  \caption{Life cycle of topics on SINA Weibo. X-axis is on a base-10 log scale}\label{fg_lifeCycle}
\end{figure}

There have been some efforts building models to predict the popularity of online contents~\cite{wu2010use}~\cite{lerman2010using}~\cite{lee2010approach}. They predicted future popularity of a given online content based on its early popularity and user's early reactions. Szabo and Huberman~\cite{szabo2010predicting} firstly found the strong linear correlation between the logarithmically transformed popularity of content at early and later times. Based on the founding, they proposed Szabo-Huberman (SH) model to predict the popularity of online content. Some researches~\cite{wu2010use} ~\cite{Pinto2013UEV} ~\cite{bao2013popularity} tried to improve the performance of SH model. There is one problem with existing methods. In Figure~\ref{fg_lifeCycle}, we can find that 74.14\% topics last for less than 10 days and 81.38\% topics last for less than 30 days. Topics change so quickly that it is important to predict the popularity of topics before they happen to earn enough time to make online advertising plans for business and so on. Existing methods focus on predicting future popularity of online contents which have already appeared based on their early popularity and users' early reactions, but they can not predict the popularity of topics which appear right now or do not even happen.

We know little about what drives popularity of topics in social network. Salganik et al.~\cite{salganik2006experimental} answered the question experimentally by measuring the impact of content quality and social influence on the popularity of cultural artifacts like songs, movies. They found that quality contributes little to their popularity, while the choices of others (social influence) affect the choice of the user and are responsibility of the success of cultural artifacts. They answer the question on an individual's point rather than group perspective. What drives the choices of a community or a network? In this paper, we try to answer the question on the perspective of the whole community of network based on user sentiment. Psychological researches~\cite{damasio2008descartes}~\cite{dolan2002emotion} show that emotions (except information) play an significant role in human decision-making. The decisions of online users to diffuse the tweets of a topic or not influence the popularity of the topic, so we consider whether the emotions of users influence topic popularity. Thelwall et al.~\cite{thelwall2011sentiment} find the close relationship between popular evens and the increase of sentiment strength. Bollen et al.~\cite{bollen2011twitter} introduced public sentiment to predict stock market. We study if the sentiment of the users on a topic is related to the popularity of the topic in a community.

In this paper, we use the number of tweets in a topic to be the popularity of the topic. We compute the sentiment of online users on key phrases of a topic in a training dataset of tweets obtained from SINA Weibo API. The sentiment of a user on the key phrases of a topic can influence the decision of discussing about it or not. For example, if a user is a big fan of movie actor Arnold Schwarzenegger, then the user is probably to discuss the topic about Arnold Schwarzenegger. We introduce Markov Random Field (MRF) model~\cite{kindermann1980markov} and graph entropy model~\cite{korner1973coding}~\cite{simonyi1995graph} to compute the total topic sentiment energy in a community based on the users' sentiment on key phrases of the topic. Pearson correlation coefficient~\cite{Wiki_Pearson} measures the linear correlation between the community sentiment energies computed from the training dataset and the real future popularity of testing topics in the testing dataset. Our experimental results show that they are linearly correlated and indicate the latent ability for predicting the popularity of topics.

Based on the finding, we propose two linear methods to predict the popularity of un-happened topics. The first one is a one variable linear regression model named LinearMRF model based on the community sentiment energy. The second is a multivariate linear regression model named EdgeMRF model, which assume that the real popularity of topics associates the energy of each edge in the graph of the community, rather than the community sentiment energy. We find that the two methods are effective in predicting the popularity of topics and the EdgeMRF model is significantly better than the LinearMRF model.

The main contributions of this paper can be summarized as follows:
\begin{itemize}
  \item We find that there is significant linear correlation between the community sentiment energy and the real popularity of topics in a community. It shows the effectiveness of sentiment in predicting the popularity of topics and the sentiment of a community can affect collective decision making of spreading a topic or not.
  \item Based on the finding, we propose two methods to predict the popularity of topics based on community user sentiment. We are the first trying to predict the popularity of topics that have not happened.
  \item Extensive experiments on two communities verify our finding and show the effectiveness of the methods in predicting the popularity of topics that have not happened.
\end{itemize}

The rest of this paper is organized as follows: Section~\ref{sc_RelatedWorks} discusses some important related works. Section~\ref{sc_framework} introduces our framework for measuring the linear correlation between the community sentiment energy and the real popularity of topics. Based on the results of the Section~\ref{sc_framework}, we introduce two models to predict the popularity of topics in Section~\ref{sc_prediction}. Finally, conclusion and discussion are provided in Section~\ref{sc_discuss}.

\section{Related Works}\label{sc_RelatedWorks}

\subsection{Popularity Prediction}

Some researches try to predict based on user sentiment. Thelwall et al.~\cite{thelwall2011sentiment} studied whether popular events are typically associated with the increase of sentiment strength. They found that popular events are normally associated with increases in negative sentiment strength. Bollen et al.~\cite{bollen2011twitter} compared the changing of Twitter mood with the Dow Jones Industrial Average (DJIA) over time and found the association relation between them. They used the relation to predict stock market.

There are a large number of works predicting popularity of online content like news, videos, and tweets in social networks. Crane and Sornette~\cite{crane2008robust} identified four main classes of the popularity evolution patterns of YouTube videos. They explained the reason of the four main classes by the combination of endogenous and exogenous effects. Szabo and Huberman~\cite{szabo2010predicting} studied the problem of predicting the popularity of online content on Digg and Youtube. They found that there were a strong linear correlation between the logarithmically transformed popularity of content at early and later times. Because of the founding, they presented a model to predict future popularity based on the popularity of early times. Pinto et al.~\cite{Pinto2013UEV} improved Szabo and Huberman's method~\cite{szabo2010predicting} using the historical information given by early population measures in Youtube. Figueiredo et al.~\cite{Figueiredo2011TOT} studied growth patterns of video popularity and the types of the referrers that most often attracted users to videos on such patterns. Wu et al.~\cite{wu2010use} proposed a reservoir computing based model to predict the near future popularity of a Web object. But their model does not outperform the Szabo-Huberman model~\cite{szabo2010predicting}, as reported by the authors. Lee et al.~\cite{lee2010approach} try to predict if a thread in a discussion forum will stat in popular in the near future using biology-inspired survival analysis method. Yin et al.~\cite{yin2012straw} tried to rank items in online sharing systems according to their excepted popularity. Their method was built on a model of user behavior and early votes of the items. Bao et al.~\cite{bao2013popularity} studied the importance of structural characteristics for the popularity of tweets on SINA Weibo. They find that the prediction accuracy can be improved by incorporating the factor of structural diversity into some existing content based methods. Gao et al.~\cite{Gao2014Effective} tried to find effective features for tweet popularity prediction. They obtain the temporal features of first 10 retweets which satisfy prediction performance. Kristina and Tad~\cite{lerman2010using} proposed a stochastic models of user behavior based method to predict popularity in social news portal Digg. They use early user reactions to a new submission to to predict how many votes the story will get a few days later. Hong et al.~\cite{hong2011predicting} cast the task of predicting the popularity of messages into a classification problem by investigating features like the content of the messages, users' social graph and so on.

There are some researches predicting hot topics in online forums or information retrieval systems, but they did not utilize the rich relations between users in social network. Tong et al.~\cite{tong2008internet} put forward an adaptive autoregressive model to predict hot topics in the near future based on the web browsing log information. Buckley~\cite{Buckley2004Topic} tried to improve the performance of information retrieval by predicting which topics work well with which approaches.

\subsection{Sentiment Analysis}

Sentiment analysis, which is also named opinion mining, is to study text with sentiments, opinions, emotions and so on. In social network platforms, users are free to express their daily sentiments and opinions. Many researches focused on this research area in social networks. Barbosa et al.~\cite{barbosa2010robust} proposed a method to detect user sentiments on Twitter tweets using characteristics of how tweets are written and meta-information of the words in the tweets. Davidov et al.~\cite{davidov2010enhanced} proposed a supervised sentiment classification framework by utilizing 50 Twitter tags and 15 emoticons. Jonathon Read~\cite{read2005using} used emoticons to reduce dependency in Machine learning methods for sentiment classification. Diakopoulos et al.~\cite{Diakopoulos2010CDP} characterized the  performance of U.S. presidential debate in 2008 by analyzing users sentiments in twitter. Bollen et al.~\cite{bollen2009modeling} extract six dimensions of mood (tension, depression, anger, vigor, fatigue, confusion) and compared the results with fluctuations recorded by stock market. Tumasjan et al.~\cite{tumasjan2010predicting} found that Twitter user sentiment validly mirror offline German federal election. Bifet et al.~\cite{bifet2010sentiment} found that Twitter messages are short and generated constantly, and well suited the characteristics of data stream. They proposed a stream based method on sentiment analysis.

Although there are many researches in computing the sentiment of a post in microblogging websites, the computing time is too long to be used, since there are 155,941,545 tweets in our ``Jackie Chan" dataset and 12,633,641 tweets in our ``Zhi-Hua Zhou" dataset. So we use the simple sentiment computing method of using emoticons like~\cite{davidov2010enhanced} and then compute the community sentiment energy of topics.

\section{Framework of our methods}~\label{sc_framework}

\subsection{Dataset and methods overview}

\begin{figure*}
  \centering
  \includegraphics[width=12cm]{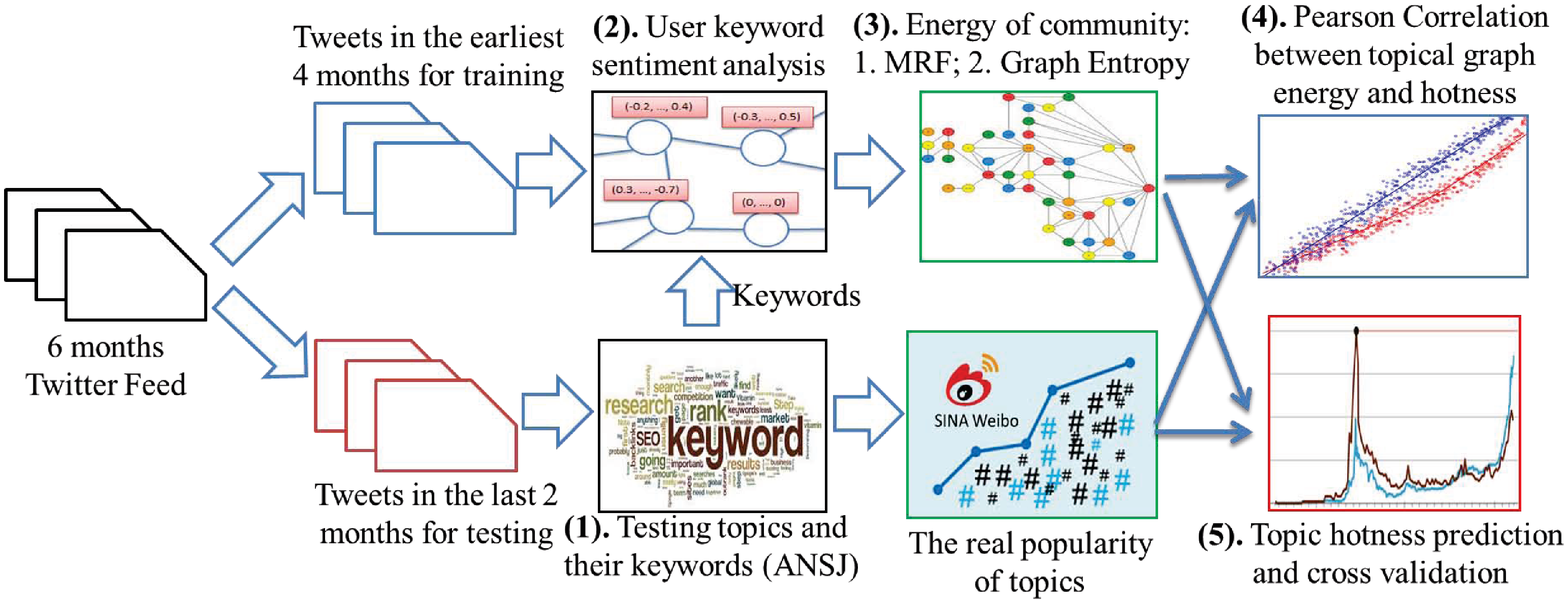}\\
  \caption{The framework of our methods. There are mainly 5 phases: (1) Extract test topics and their top-m key phrases using ANSJ toolkit; (2) Analyze user's sentiment on the top-m key phrases of topics; (3) Compute the total sentimental energy of the community using Markov Random Field (MRF) model and graph energy model; (4) Measure the linear correlation between the total sentimental energy of the community and the real popularity of topics; (5) Propose two methods to predict the popularity of topics that have not happened yet.} \label{Fig_Framework}
\end{figure*}

We obtained public tweets for six months from Jul. 1st to Dec. 31st, 2014 using SINA Weibo API. We got 6,824,948,570 public tweets of 90,388,540 users. If a user $i$ retweets from user $j$ or mentions user $j$ (@username), then there is a retweet-mention-relation between user $i$ and $j$. The relations between users were extracted from the 6,824,948,570 public tweets of 90,388,540 users. User's social graph $G$ was constructed based on the retweet-mention-relations between the users. We obtained a user community in the graph $G$ of the 90,388,540 users from a specific user who is the only one in the initial seed set $SeedSet$ based on the retweet-mention-relations. The community is made by the neighbors at distance up to $MaxDepth$. We got a community of 137,613 users from the specific user ``Jackie Chan"~\footnote{Homepage of Jackie Chan on SINA Weibo. http://weibo.com/jackiechan.}, who is a famous Hong Kong movie actor~\footnote{Jackie Chan. Wikipedia. http://en.wikipedia.org/wiki/Jackie\_Chan.}. Another community is from user ``Zhi-Hua Zhou"~\footnote{Homepage of Zhi-Hua Zhou on SINA Weibo. http://weibo.com/zhouzh2012.}, who is the ACM/AAAS/AAAI/IEEE Fellow majoring in machine learning of Nanjing University~\footnote{Zhi-Hua Zhou's Homepage. https://cs.nju.edu.cn/zhouzh/.}. There are 16,628 users in the ``Zhi-Hua Zhou" community. The maximum depth $MaxDepth$ is 4 in the ``Jackie Chan" community and 3 in the ``Zhi-Hua Zhou" community, since there are too many users which will take too long time to finish computing in the ``Zhi-Hua Zhou" community if $MaxDepth$ is 4.

We extracted public tweets of the users in the two communities for six months from Jul. 1st to Dec. 31st, 2014 on SINA Weibo. There are totally 155,941,545 tweets in the ``Jackie Chan" dataset and 12,633,641 tweets in the ``Zhi-Hua Zhou" dataset. We split the dataset into two parts: train dataset which contains 110,938,220 tweets in the ``Jackie Chan" dataset (8,039,686 tweets in the ``Zhi-Hua Zhou" dataset) in the first four months from Jul. 1th to Oct. 31 and test dataset which contains 45,003,325 tweets in the ``Jackie Chan" dataset (4,593,955 tweets in the ``Zhi-Hua Zhou" dataset) in the last two months from Nov. 1 to Dec. 31.

As shown in Figure~\ref{Fig_Framework}, our methods proceed in five phases. In the first phase, we extract testing topics from the test dataset of the last two months and the details are in Section~\ref{subsc_topics}. For each testing topic, we gather all tweets in the topic to one document and extract top-10 key phrases from the document using the key phrase extraction tool ANSJ~\footnote{ANSJ: https://github.com/NLPchina/ansj\_seg} which is an open source Chinese word processing toolkit. In the second phase, we compute the sentiment of every user on the top-10 key phrases of each testing topic from the training dataset of the first 4 months. A 10-dimensional sentiment vector of the top-10 key phrases of a topic is built for a user. Each dimension in the sentiment vector corresponds to one of the top-10 key phrases of the topic. The details of this phase are in Section~\ref{subsc_userSentiment}. In the third phase, latent energy of the user community on a topic is calculated based on the sentiment vector built in the second phase using the training dataset. Markov random field (MRF) model~\cite{kindermann1980markov} and graph entropy model~\cite{korner1973coding}~\cite{simonyi1995graph} are employed to compute the latent community sentiment energy of the user community. The details of this phase are in Section~\ref{sc_MRF} and~\ref{sc_graphEntropy}. In the fourth phase, we investigate the hypothesis that the community sentiment energy of a topic as computed based on user sentiment is predictive of future popularity of the topic. We use Pearson correlation coefficient~\footnote{Pearson correlation coefficient: https://en.wikipedia.org/wiki/Pearson\_product-moment\_correlation\_coefficient} to correlate the real testing topic popularity in the testing dataset to the community sentiment energy computed in the training dataset. The details of this phase are in Section~\ref{sc_correlation}. In the fifth phase, we propose two methods to predict the popularity of the topics and employ the real popularity of the topics to validate the effectiveness of our methods. The details of this phase are in Section~\ref{sc_prediction}.

\subsection{Testing topics and their key phrases}~\label{subsc_topics}

On SINA Weibo, users can annotate tweets with hashtags (``\#hashtag\#" on SINA Weibo, ``\#hashtag" on Twitter) to indicate the ongoing topics. For example, ``\#Rio2016\#" (``\#Rio2016" on Twitter) is the hashtag indicating the topic of``Summer Olympics 2016". We use hashtags to represent topics and use the hashtags to extract tweets of topics from the test dataset (if a tweet contains a hashtag, then the tweet is about the topic of the hashtag). It is a convenient way to categorize tweets into different topics. In this paper, we use the number of tweets in a topic to be the popularity of the topic. We count the occurrence number of every hashtag in the tweets of our test datasets to be the popularity of the corresponding topic. If a tweet has multiple hashtags, then the tweet will belong to multiple topics in our experiments.

\begin{figure}[t]
\centering
\subfigure[``Jackie Chan" dataset]{
   \includegraphics[width=5.8cm] {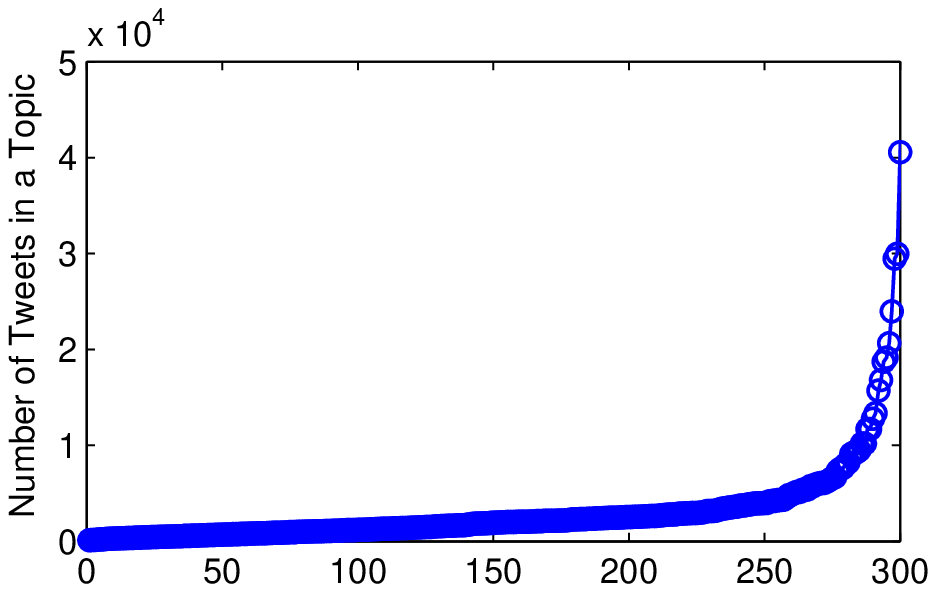}
   \label{NumOfTweets}
 }
 \subfigure[``Zhi-Hua Zhou" dataset]{
   \includegraphics[width=5.8cm] {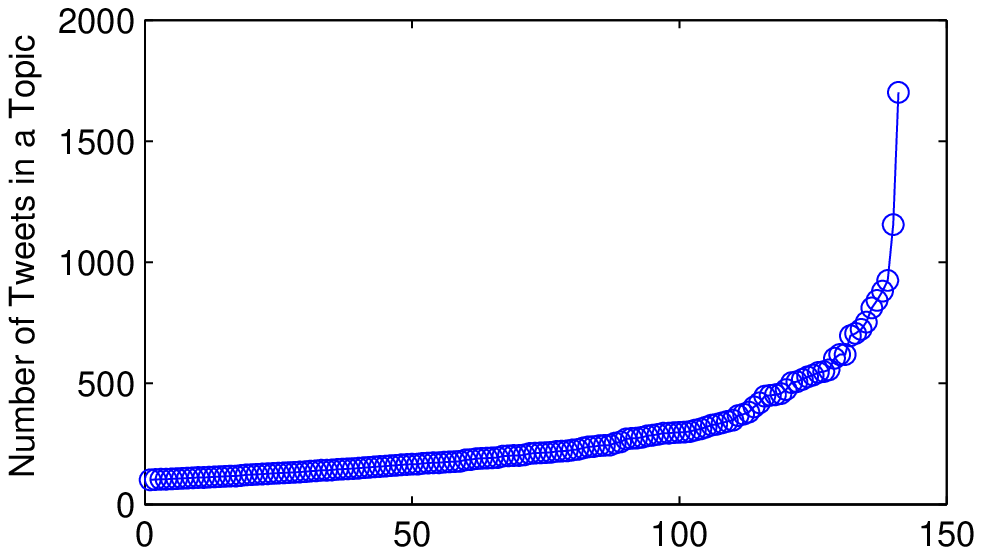}
   \label{NumOfTweetsZhou}
 }
\caption{Number of tweets in the testing topics of the two communities}
\label{fig_NumOfTweets}
\end{figure}

To make sure that the topics have fully diffused and the popularity of the topics is correctly calculated, we only extracted topics (hashtags) which begin in the first month (November 2014) of the test dataset. Topics which begin in the second month (December 2014) may have not fully diffused and the number of tweets in the topics may not be correct. To count the number of tweets in the topics, the whole tweets in the test dataset of the two months (November and December, 2014) are used. We remove unpopular topics which are less than 100 tweets and get 5150 topics in the test dataset of the ``Jackie Chan" community and 224 topics in the test dataset of the ``Zhi-Hua Zhou" community. After removing topics with same number of tweets, we finally get 298 topics in the ``Jackie Chan" community and 141 topics in the ``Zhi-Hua Zhou" community for testing. Figure~\ref{fig_NumOfTweets} shows the real popularity (the number of tweets) of the 298 topics and the 141 topics for testing.

ANSJ is an open source Chinese word processing toolkit. It can be used for keyword extraction, word segmentation and so on. NLPIR (ICTCLAS)~\footnote{NLPIR: http://ictclas.nlpir.org/} is also a popular Chinese word processing tool that can extract key phrases from a document, but NLPIR (ICTCLAS) can not handle large document quickly, so we choose ANSJ to process key phrases extraction. We add stopwords to ANSJ toolkit to avoid the appearance of them in the results. We gather all tweets in a topic to one document and then extract the top-m key phrases from the document using ANSJ. A topic $k$ is then represented as a bag of key phrases as shown in Equation~\ref{eq_topic}.
\begin{equation}\label{eq_topic}
{Topic_k=[term_{k,1}, term_{k,2},\cdots,term_{k,m}]}
\end{equation}
where $term_{k,n} (n=1,2,\cdots,m)$ is one of the top-m key phrases extracted from topic $k$ using ANSJ. We set m to be 10 empirically in all our experiments.

\subsection{User sentiment on a topic}~\label{subsc_userSentiment}

Since a topic is represented as a bag of key phrases as shown in Equation~\ref{eq_topic}, then the sentiment of a user on a topic can be represented by the sentiment on the top-m key phrases. We compute the user sentiment of the testing topics from the training dataset. Suppose $s(u_i)_{k,n}$ is the sentiment of user $u_i$ on key phrase $term_{k,n} (n=1,2,\cdots,m)$, then the sentiment $sen(u_i,k)$ of user $u_i$ on topic $k$ is represented as Equation~\ref{eq_sentimentTopic}.
\begin{equation}\label{eq_sentimentTopic}
sen(i,k) = [s(u_i)_{k,1}, s(u_i)_{k,2},\cdots, s(u_i)_{k,m}]
\end{equation}

As the length of a tweet on SINA Weibo is very short without exceeding 140 characters, we simply assume that the sentiment of user $u_i$ on a key phrase of a tweet is equal to the sentiment of user $u_i$ on the tweet which contains the key phrase. If the key phrase does not exist in the tweet, the sentiment of $u_i$ on the tweet will be 0. There are a large percent of tweets that do not contain the key phrases of a topic, but we use months of user tweets in the training dataset and we compute sentiment energy of a community of users rather than a user. So there are still many tweets containing the key phrases of a topic for computing community sentiment energy.
\begin{figure}[htbp]
  \centering
  \includegraphics[width=6.0cm]{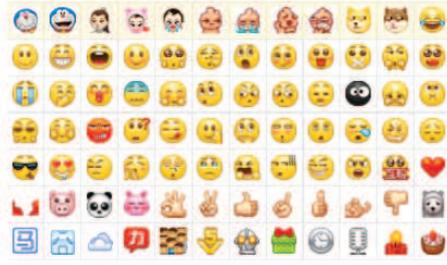}\\
  \caption{Typical emoticons on SINA Weibo}\label{fig_emotion}
\end{figure}

In microblogging websites, the words in tweets are usually non-standardized and personalized. There are always a lot of new words existing and traditional methods for computing sentiment is hard to be used. There are so many tweets (110,938,220 tweets in the train dataset) that the amount of calculation is very huge. Many researches in computing the sentiment of a post in microblogging websites can not be used due to the computational speed in so many tweets. Emoticons are convenient ways for users in microblogging website to show their emotions and sentiment when they are posting tweets. We find that there are 52,602,698 tweets contains emoticons in the 110,938,220 tweets of the ``Jackie Chan" train dataset. It is 47.42\% of all tweets in the train dataset. We compute the user sentiment on a tweet based on 436 emoticons on SINA Weibo as shown in Figure~\ref{fig_emotion} like~\cite{davidov2010enhanced}. We annotate the 436 emoticons to be positive, negative or neutral manually. There are three volunteers to annotate the 436 emoticons. If two or all of the three volunteers think the emoticon is positive (negative or neutral), then it is. There are no cases that all the three volunteers hold different opinions in the process of annotation since the emoticons are relatively easy to annotate. Suppose there are $pos$ positive emoticons and $neg$ negative emoticons on a tweet $j$ which contains the key phrase $term_{k,n} (n=1,2,\cdots,m)$, then the sentiment $stweet(u_i,j)$ of the user $u_i$ on the tweet $j$ and the sentiment $sphrase(u_i,j)_{k,n}$ of the user $u_i$ on the key phrase $term_{k,n} (n=1,2,\cdots,10)$ of tweet $j$ can be calculated as Equation~\ref{eq_sentweet}.
\begin{equation}\label{eq_sentweet}
%\begin{split}
 sphrase(u_i,j)_{k,n} = stweet(u_i,j)= (pos-neg)/(pos+neg)
%\end{split}
\end{equation}

If key phrase $term_{k,n} (n=1,2,\cdots,m)$ does not appear in tweet $j$, then $sphrase(u_i,j)_{k,n}=0$. If $Tweet(u_i)$ is the set of all tweets of user $u_i$, then the sentiment $sen(u_i)_{k,n}$ of user $u_i$ on key phrase $term_{k,n} (n=1,2,\cdots,m)$ in Equation~\ref{eq_sentimentTopic} can be calculated as shown in Equation~\ref{eq_sen}.
\begin{equation}\label{eq_sen}
s(u_i)_{k,n} = \frac{1}{|Tweet(u_i)|}\sum\limits_{j \in Tweet(u_i)} {{\rm{sphrase(u_i,j}}{{\rm{)}}_{k,n}}}
\end{equation}

\subsection{Community Sentiment Energy based on Markov Random Field model}\label{sc_MRF}

Markov Random Field (MRF)~\cite{kindermann1980markov}, which is also known as Markov network or undirected graphical model, is a set of random variables having a Markov property in an undirected graph. MRF is commonly used in statistical machine learning to model joint distributions, such as modeling image de-noising~\cite{li2009markov}, information retrieval~\cite{metzler2005markov} and so on. A MRF is an undirected graph $G=(\mathcal{V},\mathcal{E})$, where $\mathcal{V} =\{u_1, u_2,..., u_N\}$ is the set of users in social network, $\mathcal{E}$ is the set of retweet-mention-relations between users. Each node $u_i (i=1,2,\cdots,N)$ is associated with a random variable (RV), which is the sentimental vector $sen(u_i,k) (i=1, 2,..., N)$ of user $u_i$ on topic $k$. To satisfy the Markov property, we assume that \emph{information only diffuses to neighbours}. Then MRF satisfies $p(sen(u_i,k)|{\{sen(u_j,k)\}}_{j=1, 2,..., N})=p(sen(u_i,k)|{\{sen(u_j,k)\}}_{u_j\in {\mathcal{N}_i}})$, where ${\mathcal{N}_i}$ is the neighbourhood of node $u_i$, $u_j\in {\mathcal{N}_i}$ if and only if $(u_i, u_j)\in \mathcal{E}$.

According to MRF, suppose $C_x$ is one of the max cliques in the graph and $C=\bigcup\limits_x C_x$, the total energy of graph $G$ on topic $k$ is $E(G,k)=E(C,k)=\sum\limits_x {E({C_x},k)} $. To simplify handling, we only consider the cliques with two nodes. In other words, we only consider the cliques where two nodes are connected by an edge. The energy of a max clique with multi nodes is simply the sum of energy of each sub-clique with two nodes. We define two kinds of energy function of cliques with two nodes: \textbf{cosine measure} and \textbf{average length}. We try to compute the community sentiment energy between $u_i$ and $u_j$ using the two functions to evaluate the probability of $u_i$ and $u_j$ talking about a topic in microblogging websites. We find that if two users have the same sentiment on a topic, they like to talk about it and the sentimental energy between them will be large. If they have opposite sentiment on a topic, they also like to debate on it and the sentimental energy between them will be large too. So traditional Euclidean Distance does not suit this case, since the sentimental energy will be zero if two users have exactly the same sentiment on a topic. So we use the average length of the two sentimental vectors to the zero-vector to represent the sentimental energy between two users. The absolute value of cosine similarity (\textbf{cosine measure}) of two vectors also suits this case.

For energy function \textbf{cosine measure}, Equation~\ref{Eq_cosine} is the energy of the clique $C_{ij}$ between two nodes $u_i$ and $u_j$. It is the absolute value of cosine similarity between sentimental vectors $sen(u_i,k) (i=1, 2,..., N)$ and $sen(u_j,k) (j=1, 2,..., N)$ of node $u_i$ and $u_j$ on topic $k$.
\begin{equation}\label{Eq_cosine}
%\begin{split}
E({C_{ij}},k) = E({sen(u_i,k)},{sen(u_j,k)}) \\
= \frac{{\left| {{sen(u_i,k)} \bullet {sen(u_j,k)}} \right|}}{{\left| {{sen(u_i,k)}} \right| \bullet \left| {{sen(u_j,k)}} \right|}}
%\end{split}
\end{equation}

For energy function \textbf{average length}, Equation~\ref{Eq_length} is the energy of the clique $C_{ij}$ between two nodes $u_i$ and $u_j$. It is the average length of sentiment vectors $sen(u_i,k) (i=1, 2,..., N)$ and $sen(u_j,k) (i=1, 2,..., N)$ of node $u_i$ and $u_j$.
\begin{equation}\label{Eq_length}
%\begin{split}
E({C_{ij}},k) = E({sen(u_i,k)},{sen(u_j,k)}) \\
= \frac{{\left| {{sen(u_i,k)}} \right| + \left| {{sen(u_j,k)}} \right|}}{2}
%\end{split}
\end{equation}

The community sentiment energy of graph (community) $G$ on topic $k$ can be calculated as Equation~\ref{Eq_energy},
\begin{equation}\label{Eq_energy}
%\begin{split}
  E(G,k) = E(U,k) = \sum\limits_C {E({C_{ij}},k)} \\
  = \sum\limits_{( {u_i},{u_j})  \in \mathcal{E}} {E({sen(u_i,k)},{sen(u_j,k)})}
%  \end{split}
\end{equation}
where $\mathcal{E}$ is the edge set of Graph $G$.

\subsection{Community Sentiment Energy based on Graph Entropy Model}\label{sc_graphEntropy}

Shannon~\cite{shannon2001mathematical} proposed entropy to quantify information in information theory. Shannon's measure of information is the number of bits to represent the amount of uncertainty (randomness) in a data source. K{\"o}rner et al. ~\cite{korner1973coding}~\cite{simonyi1995graph} introduced graph entropy associated with the graph based on the Shannon entropy. Anand and Bianconi~\cite{anand2009entropy} studied different definition of entropy of network ensembles for the quantification of the complexity of networks. Cruz et al.~\cite{cruz2011entropy} proposed a method to detect community based on the graph entropy. In this paper, we compute the graph entropy of a community (energy of a community) to show the quantification of the complexity of the community. Graph entropy, which measures the latent community sentiment energy of the graph, is computed based on the user sentiment vector of a topic. Given a graph $G$, suppose $p_{eij}$ is the probability of edge $eij$ of user $u_i$ and $u_j$ to communicate about the topic $k$. Like paper \cite{anand2009entropy} and \cite{cruz2011entropy}, the graph entropy (the sentiment energy) of the community can be calculated as Equation~\ref{eq_entropy}.
\begin{equation}\label{eq_entropy}
\begin{split}
E(G,k) =& -\sum\limits_{eij\in \mathcal{E}} (p_{eij}\log(p_{eij}) + (1-p_{eij})\log (1-p_{eij})) \\
=& -\sum\limits_C ({E({C_{ij}},k)}\log({E({C_{ij}},k)}) + (1-{E({C_{ij}},k)})\log (1-{E({C_{ij}},k)}))
\end{split}
\end{equation}
where $\mathcal{E}$ is the edge set of Graph $G$. The probability $p_{eij}$ of edge $eij$ of user $u_i$ and $u_j$ to communicate about the topic $k$ can be computed based on the user sentiment vectors $sen(u_i,k) (i=1, 2,..., N)$ and $sen(u_j,k) (j=1, 2,..., N)$. The function \textbf{cosine measure} in Equation~\ref{Eq_cosine} and \textbf{average length} in Equation~\ref{Eq_length} are used to compute $p_{eij}$. Suppose $sen(u_i,k)$ and $sen(u_j,k)$ are the sentiment vectors of user $u_i$ and $u_j$ for topic $k$, for function \textbf{cosine measure}, we can calculate $p_{eij}$ as $p_{eij}= E({C_{ij}},k)=\frac{{\left| {{sen(u_i,k)} \bullet {sen(u_j,k)}} \right|}}{{\left| {{sen(u_i,k)}} \right| \bullet \left| {{sen(u_j,k)}} \right|}}$; for function \textbf{average length}, we can calculate $p_{eij}$ as $p_{eij}= E({C_{ij}},k)= \frac{{\left| {{sen(u_i,k)}} \right| + \left| {{sen(u_j,k)}} \right|}}{2}$.

\subsection{Linear Correlation Between Community sentiment Energy and the Real Popularity of Topics} \label{sc_correlation}

Section~\ref{sc_MRF} and~\ref{sc_graphEntropy} shows two different models of MRF and graph entropy in computing community sentiment energy on a topic. Two energy functions \textbf{cosine measure} in Equation~\ref{Eq_cosine} and \textbf{average length} in Equation~\ref{Eq_length} are used to compute the energy or the probability $p_{eij}$ to communicate between user $u_i$ and user $u_j$. In this section, we study whether the community sentiment energy is predictive of the real popularity of a topic in the community. We study the linear correlation between the graph energy of a community and the real popularity. In this paper, we define the number of tweets in a topic to be the real popularity of the topic.

Pearson correlation coefficient~\cite{Wiki_Pearson} developed by Karl Pearson is widely used to measure the linear correlation (dependence) between two variables. Correlation coefficient $r$ is to measure the strength of the relationship between the two variables. In general, the higher the correlation coefficient $r$, the stronger the relationship. Significance tests are done to test if correlation coefficient is significantly different from zero. There are often three typical significance level of p-value $p$: 0.01, 0.05 and 0.1. If p-value $p$ is smaller than a significance level such as 0.05, then correlation coefficient is significantly different from zero in the significance level 0.05. We compute the $r$ and $p$ to test the linear correlation between the community sentiment energy and the real popularity.

\begin{figure}[t]
\centering
\subfigure[``Jackie Chan" dataset]{
   \includegraphics[width=5.6cm] {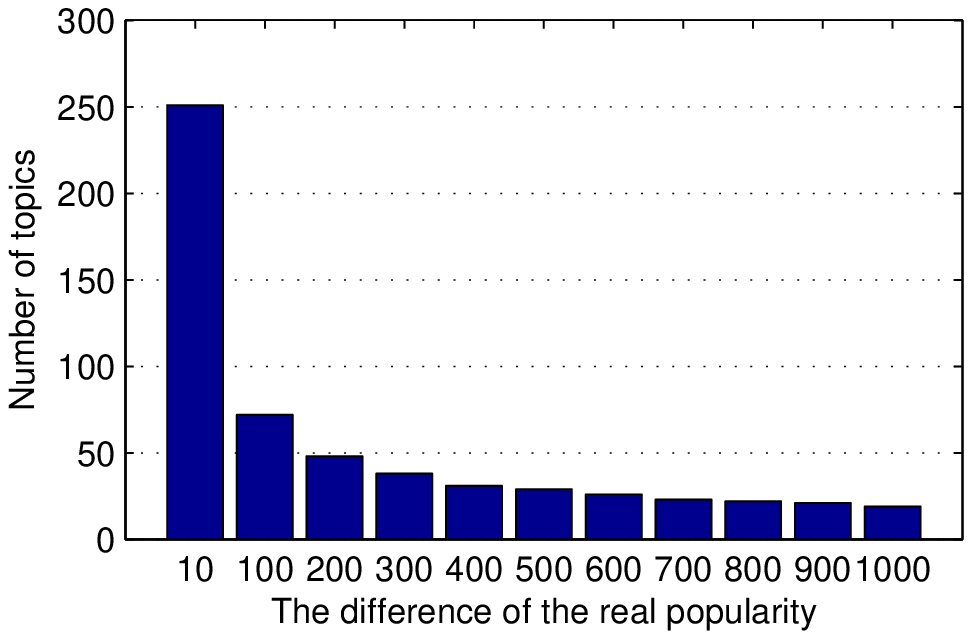}
   \label{NumOfDatasets}
 }
 \subfigure[``Zhi-Hua Zhou" dataset]{
   \includegraphics[width=5.8cm] {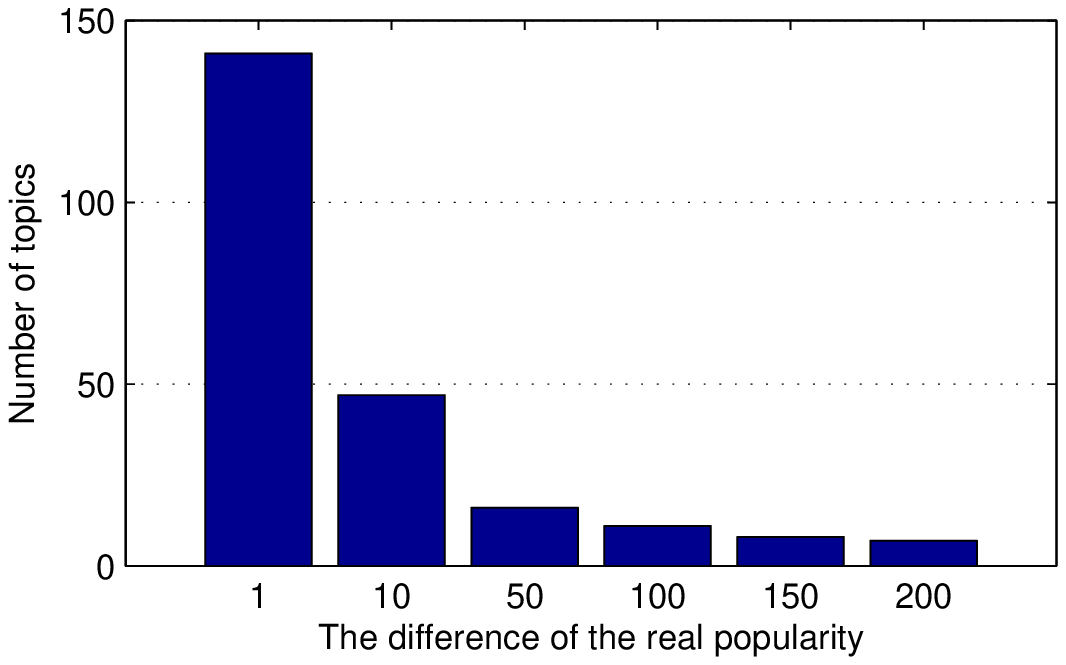}
   \label{NumOfDatasetsZhou}
 }
\caption{The number of topics in the 11 datasets. The gaps of the real popularity between every two topics in the test datasets are shown on the X axis}
\label{fig_NumOfTheDatasets}
\end{figure}

To test the linear relations between the community sentiment energy and the real popularity of topics, we constructed 11 test datasets (6 test datasets) according to the gap of the number of tweets between every two topics in the ``Jackie Chan" (``Zhi-hua Zhou") dataset. The gap of the real popularity between every two topics in a test dataset are equal or larger than $Num (Num = 10, 100, 200, 300, 400, 500, 600, 700, 800, 900, 1000)$ in the ``Jackie Chan" dataset and $Num (Num =1, 10, 50, 100, 150, 200)$ in ``Zhi-hua Zhou" dataset. For example, for dataset $Num=10$, The gap of the real popularity between every two topics are equal or larger than 10. Figure~\ref{fig_NumOfTheDatasets} shows the number of topics in the test datasets of the two communities.

We compute the Pearson correlation coefficients of the community sentiment energy and the real popularity with the change of the different $Num$ in the test dataset of the two communities. Table~\ref{tab:Pearson} and~\ref{tab:Pearson_zhou} shows the pearson correlation coefficient results of community sentiment energy and real popularity on the 11 test datasets of ``Jackie Chan" community and 5 test datasets of ``Zhi-hua Zhou" community. According to~\cite{dancey2007statistics}, the strength of correlation can be classified to five categorisations: Zero(0), Weak(0.1-0.3), Moderate(0.4-0.6), Strong(0.7-0.9) and Perfect(1.0). So if the p-value is smaller than the significance level and the Pearson correlation coefficient r-value is larger than 0.4, the linear correlation between the topic sentiment energy of a community and the real popularity is significant.

In Table~\ref{tab:Pearson} and~\ref{tab:Pearson_zhou}, we can find that the community sentiment energy computed by MRF model with energy function ``cosine measure" has a significant linear correlation with the real topic popularity if the different $Num$ is equal or larger than 100 in the ``Jackie Chan" dataset and 1 in the ``Zhi-hua Zhou" dataset. It means that it's possible to predict the popularity of topics that did not happen with a small error. Significant test student's t-test~\footnote{Student's t-test. Wikipedia. https://en.wikipedia.org/wiki/Student's\_t-test} is used to indicate statistically significant improvements of one method over another methods. The p-values of student's t-test of the method ``MRF+CosineMeasure" over the other three methods ``GraphEntropy+AverageLength", ``MRF+AverageLength", ``GraphEntropy+CosineMeasure" are 7.5e-007, 0.0011 and 1.5e-004 in the ``Jackie Chan" dataset. The p-values of t-test over the three methods in the ``Zhi-hua Zhou" dataset are 0.0022, 0.0017 and 0.0021.  They are all smaller than typical significant level 0.01. It means that the method ``MRF+CosineMeasure" is the best one of the four methods in computing the community sentiment energy. The method ``MRF+AverageLength" in ``Jackie Chan" dataset is also better than ``GraphEntropy+AverageLength" and ``GraphEntropy+CosineMeasure", since the p-values of ``MRF+AverageLength" over ``GraphEntropy+AverageLength" and ``GraphEntropy+CosineMeasure" are 1.1e-004 and 8.5e-004, which are smaller than significant level 0.01. But in the ``Zhi-hua Zhou" dataset, the p-value of method ``MRF+AverageLength" and ``GraphEntropy+CosineMeasure" is 0.1215, so no one is significant better than the other one.

\begin{table*}[t]
  \centering
  \caption{Pearson correlation coefficient results of community sentiment energy and real popularity in ``Jackie Chan" dataset}
    \begin{tabular}{c|cc|cc|cc|cc}
    \hline
    \multirow{3}[4]{*}{Datasets} & \multicolumn{2}{|c}{GraphEntropy} & \multicolumn{2}{|c}{MRF} & \multicolumn{2}{|c}{MRF} & \multicolumn{2}{|c}{GraphEntropy} \\
          & \multicolumn{2}{|c}{+AverageLength} & \multicolumn{2}{|c}{+ AverageLength} & \multicolumn{2}{|c}{+CosineMeasure} & \multicolumn{2}{|c}{+ CosineMeasure} \\
          \cline{2-9}
          & r     & p     & r     & p     & r     & p     & r     & p \\
    \hline
    10  & 0.1561 & 0.0128 & 0.2065 & 9.50E-04 & 0.2832 & 5.10E-06 & 0.1492 & 0.0169 \\
    100 & 0.1572 & 0.1971 & 0.3329 & 0.0046 & \textbf{0.4295} & \textbf{1.60E-04$\star\star\star$} & 0.1929 & 0.1122 \\
    200 & 0.1064 & 0.4764 & 0.3935 & 0.0062 & \textbf{0.5055} & \textbf{2.40E-04$\star\star\star$} & 0.1884 & 0.2048 \\
    300 & 0.2225 & 0.1793 & \textbf{0.4067} & \textbf{0.0113$\star\star$} & \textbf{0.4157} & \textbf{0.0094$\star\star\star$} & 0.2399 & 0.1468 \\
    400 & 0.2611 & 0.156 & \textbf{0.5079} & \textbf{0.003$\star\star\star$} & \textbf{0.5954} & \textbf{4.10E-04$\star\star\star$} & 0.3201 & 0.0792 \\
    500 & 0.2895 & 0.1277 & \textbf{0.4459} & \textbf{0.0153$\star\star$} & \textbf{0.5580} & \textbf{0.0017$\star\star\star$} & 0.1529 & 0.4286 \\
    600 & 0.2769 & 0.1709 & \textbf{0.5643} & \textbf{0.0041$\star\star\star$} & \textbf{0.6384} & \textbf{4.40E-04$\star\star\star$} & 0.256 & 0.2273 \\
    700 & 0.2207 & 0.3115 & \textbf{0.5485} & \textbf{0.0067$\star\star\star$} & \textbf{0.5111} & \textbf{0.0127$\star\star$} & 0.0876 & 0.6911 \\
    800 & 0.2767 & 0.2125 & \textbf{0.4853} & \textbf{0.022$\star\star$} & \textbf{0.5339} & \textbf{0.0105$\star\star$} & 0.3459 & 0.1149 \\
    900 & \textbf{0.4073} & \textbf{0.0668$\star$} & 0.3973 & 0.0828 & \textbf{0.5792} & \textbf{0.0059$\star\star\star$} & 0.1128 & 0.6264 \\
    1000 & 0.3966 & 0.0928 & \textbf{0.5509} & \textbf{0.0118$\star\star$} & \textbf{0.6442} & \textbf{0.0029$\star\star\star$} & -0.1993 & 0.3995 \\
    \hline
    \end{tabular}%
  \label{tab:Pearson}%
  \\( Significance tests p-value$<$0.01: $\star\star\star$,   p-value$<$0.05: $\star\star$,   p-value$<$0.1: $\star$)
\end{table*}%

\begin{table*}[t]
  \centering
  \caption{Pearson correlation coefficient results of community sentiment energy and real popularity in ``Zhi-hua Zhou" dataset}
    \begin{tabular}{c|cc|cc|cc|cc}
    \hline
    \multirow{3}[4]{*}{Datasets} & \multicolumn{2}{|c}{GraphEntropy} & \multicolumn{2}{|c}{MRF} & \multicolumn{2}{|c}{MRF} & \multicolumn{2}{|c}{GraphEntropy} \\
          & \multicolumn{2}{|c}{+AverageLength} & \multicolumn{2}{|c}{+ AverageLength} & \multicolumn{2}{|c}{+CosineMeasure} & \multicolumn{2}{|c}{+ CosineMeasure} \\
          \cline{2-9}
          & r     & p     & r     & p     & r     & p     & r     & p \\
    \hline
    1  & 0.2107 & 0.0125 & 0.1709 & 0.0435 & \textbf{0.4196} & \textbf{2.4682e-7$\star\star\star$} & 0.3758 & 4.7416e-6 \\
    10 & \textbf{0.4201} & \textbf{0.0033$\star\star\star$} & 0.3928 & 0.0063 & \textbf{0.6153} & \textbf{4.1738e-6$\star\star\star$} & \textbf{0.4307} & \textbf{0.0025$\star\star\star$} \\
    50 & \textbf{0.6697} & \textbf{0.0045$\star\star\star$} & \textbf{0.6568} & \textbf{0.0057$\star\star\star$} & \textbf{0.7275} & \textbf{0.0014$\star\star\star$} & \textbf{0.5636} & \textbf{0.0230$\star\star$} \\
    100 & \textbf{0.6061} & \textbf{0.0481$\star\star$} & \textbf{0.5656} & \textbf{0.0698$\star$} & \textbf{0.8623} & \textbf{6.3824e-4$\star\star\star$} & \textbf{0.6286} & \textbf{0.0383$\star\star$} \\
    150 & \textbf{0.6875} & \textbf{0.0595$\star$} & 0.6201 & 0.1010 & \textbf{0.8488} & \textbf{0.0077$\star\star\star$} & \textbf{0.7162} & \textbf{0.0457$\star\star$} \\
    200 & 0.6034 & 0.1514 & 0.5506 & 0.2003 & \textbf{0.9008} & \textbf{0.0056$\star\star\star$} & \textbf{0.7539} & \textbf{0.0503$\star$} \\
    \hline
    \end{tabular}%
  \label{tab:Pearson_zhou}%
  \\( Significance tests p-value$<$0.01: $\star\star\star$,   p-value$<$0.05: $\star\star$,   p-value$<$0.1: $\star$)
\end{table*}%

We try different splits of training dataset and testing dataset to show the effect of the method ``MRF+CosineMeasure" which is the best one of the four proposed methods. In previous experiments, we use 4 months' tweets to be training dataset for computing community sentiment energy, the other two months' tweets to be the testing dataset. We try to use early 3 months' (2 months', 1 month) tweets to be the training dataset and the other tweets to be the testing dataset. Table~\ref{tab:split_zhou} shows the experimental results of different splits of training dataset and testing dataset. We can find that all r-values of Pearson correlation coefficient are larger than 0.4. The p-values of student t-test of 4 months' tweets training to 3 months' (2 months', 1 month) tweets training are 0.5471, 0.3294, 0.1047. The p-values are all not less than 0.1 and the values show that 4 months training dataset is not significantly better than the 3 months' (2 months', 1 month) training dataset. But the p-values are decreasing with the decrease of the number of tweets in the training dataset. So we can find that the decreasing of the number of tweets in the training dataset does not significantly degrade the performance. It means we can use less training tweets to predict the popularity of topics.

\begin{table*}[t]
  \centering
  \caption{ Pearson correlation coefficient results of community sentiment energy and real popularity of ``MRF+CosineMeasure" method in different splits of ``Zhi-hua Zhou" dataset}
    \begin{tabular}{c|cc|cc|cc|cc}
    \hline
    \multirow{2}[4]{*}{Datasets} & \multicolumn{2}{|c}{1 months training} & \multicolumn{2}{|c}{2 months training} & \multicolumn{2}{|c}{3 months training} & \multicolumn{2}{|c}{4 months training} \\
          \cline{2-9}
          & r     & p     & r     & p     & r     & p     & r     & p \\
    \hline
    1  & \textbf{0.4769} & \textbf{2.5738e-9} & 0.4690 & 5.0831e-9 & 0.4380 & 6.2311e-8 & 0.4196 & 2.4682e-7 \\
    10 & 0.4800 & 6.3906e-4 & 0.5611 & 4.0823e-5 & 0.5936 & 1.7265e-5 & \textbf{0.6153} & \textbf{4.1738e-6} \\
    50 & 0.4722 & 0.0648 & 0.5905 & 0.0160 & 0.6872 & 0.0033 & \textbf{0.7275} & \textbf{0.0014} \\
    100 & 0.7830 & 0.0044 & 0.8607 & 6.6926e-4 & \textbf{0.8711} & \textbf{4.8043e-4} & 0.8623 & 6.3824e-4 \\
    150 & 0.7709 & 0.0251 & 0.8334 & 0.0102 & 0.8457 & 0.0082 & \textbf{0.8488} & \textbf{0.0077} \\
    200 & 0.8789 & 0.0092 & 0.8930 & 0.0068 & \textbf{0.9045} & \textbf{0.0051} & 0.9008 & 0.0056 \\
    \hline
    \end{tabular}%
  \label{tab:split_zhou}%
\end{table*}%

\section{Prediction Model and Experiments}\label{sc_prediction}

Section~\ref{sc_correlation} studied the linear correlation relation between the community sentiment energy and the real popularity in a community. The experiments show that the community sentiment energy computed based on Markov Random Field (MRF) model with energy function ``cosine measure" is the best one of the proposed methods in predicting the popularity of topics that have not appeared. In this section, we proposed two linear methods to predict the popularity of topics based on the method: \textbf{LinearMRF} and \textbf{EdgeMRF}.

\textbf{LinearMRF}. This method is based on the proved hypothesis that the community sentiment energy based on the MRF model with energy function ``cosine measure" has a significant linear correlation with the real topic popularity. We can predict the popularity of topic $k$ using Equation~\ref{eq_linearMRF},
\begin{equation}\label{eq_linearMRF}
P(G,k)=\alpha \bullet E(G,k) +\beta
\end{equation}
where $\alpha$ and $\beta$ are the parameters that need to learn. The definition of E(G,k) shows in Equation~\ref{Eq_energy}.

\textbf{EdgeMRF}. This method assumes that each edge $eij$ in the graph $G$ has different weight $\omega_{eij}$. The predicted popularity of topic $k$ has a linear correlation with the energy of each edge in graph $G$ of the community. Equation~\ref{eq_edgeMRF} shows how to compute the sentiment energy of topic $k$,
\begin{equation}\label{eq_edgeMRF}
P(G,k)= \sum\limits_{( {u_i},{u_j})  \in \mathcal{E}} {\omega_{eij} \bullet E({sen(u_i,k)},{sen(u_j,k)})} + \rho
\end{equation}
where $E({sen(u_i,k)},{sen(u_j,k)})$ is defined in Equation~\ref{Eq_cosine} and $\rho$ is a parameter which needs to learn from training.

Loss function based on ``mean squared error" estimator is constructed to learn the best parameters for predicting popularity of topics in Equation~\ref{eq_linearMRF} and~\ref{eq_edgeMRF}. Let $ST$ be the set of all topics in the train dataset, then the loss function is shown in Equation~\ref{eq_loss},
\begin{equation}\label{eq_loss}
l=\frac{1}{2|ST|} \sum \limits_{k\in ST} {(P(G,k)-real(G,k))^2}
\end{equation}
where $real(G,k)$ is the real popularity of graph $G$ on topic $k$.

Stochastic gradient descent method is introduced to minimize the total loss through training. The steps of stochastic gradient descent can be presented as follows: (1) Choose the initial values of the parameters and learning rate. (2) Repeat step (2.1) and (2.2) until an approximate minimum of the loss function is obtained: (2.1) Compute the gradient of each parameter; (2.2) Update the values of the arguments toward the descending gradient.

We take parameter $\omega_{eij}$ for example to illustrate how we use stochastic gradient descent to learn parameters in our framework. The gradient of parameter $\omega_{eij}$ can be calculated as Equation~\ref{eq_gradient},
\begin{equation}\label{eq_gradient}
 \frac{\partial l}{\partial \omega_{eij}}= \frac{1}{|ST|}\sum \limits_{k\in ST} {(P(G,k)-real(G,k))}\bullet E({C_{ij}},k)
\end{equation}
where $E({C_{ij}},k)$ is computed in Equation~\ref{Eq_cosine}. Parameter $\omega_{eij}$ can be updated according to Equation~\ref{eq_weij},
\begin{equation}\label{eq_weij}
{\omega_{eij}}^{(t+1)}={\omega_{eij}}^{(t)}-\eta \bullet \frac{\partial l}{\partial \omega_{eij}}
\end{equation}
where $\eta$ is the learning rate and $\frac{\partial l}{\partial \omega_{eij}}$ is calculated in Equation~\ref{eq_gradient}.

\begin{figure}[t]
\centering
\subfigure[$Num=800$ of ``Jackie Chan" dataset ($R^2=0.3943$)]{
   \includegraphics[width=5.8cm] {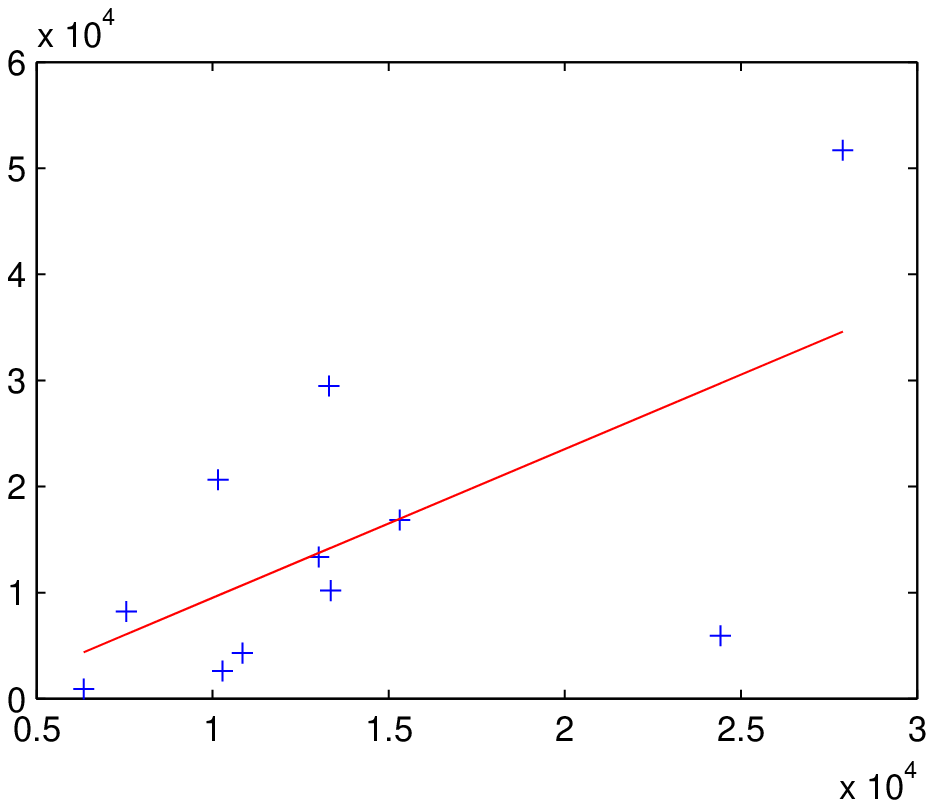}
   \label{regressionChen}
 }
 \subfigure[$Num=200$ of ``Zhi-Hua Zhou" dataset ($R^2=0.8115$)]{
   \includegraphics[width=5.8cm] {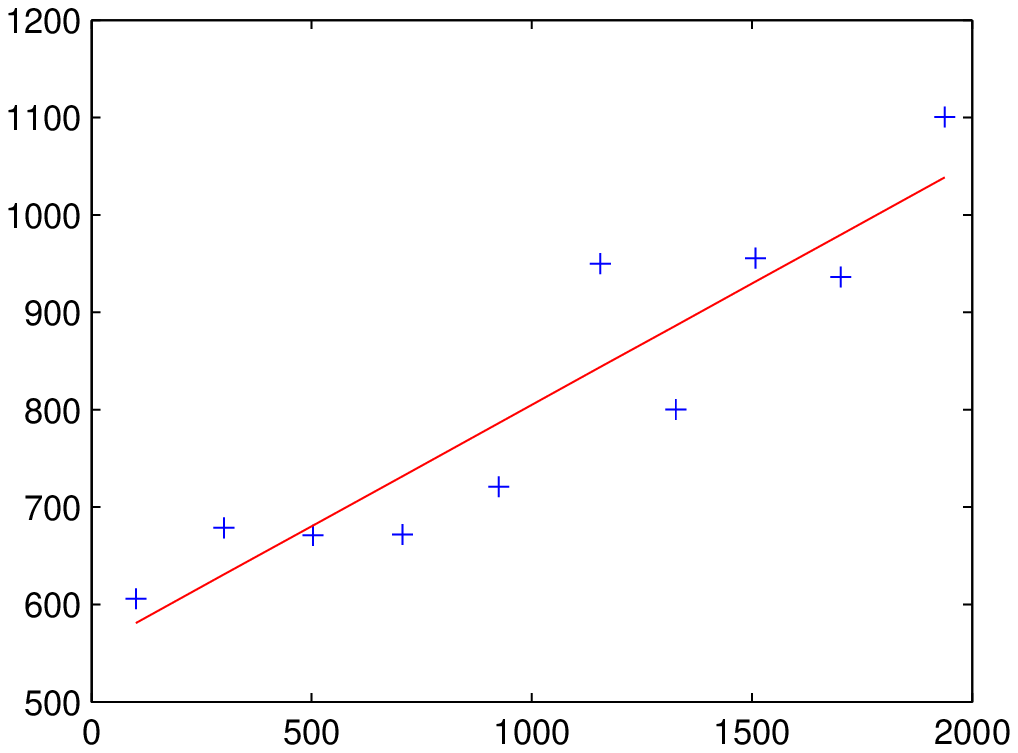}
   \label{regressionZhou}
 }
\caption{Regression Line of the LinearMRF method in the two communities}
\label{fig_regression}
\end{figure}
Figure~\ref{fig_regression} shows the regression lines of the LinearMRF method. We use coefficient of determination~\footnote{Coefficient of determination. Wikipedia. https://en.wikipedia.org/wiki/Coefficient\_of\_determination} to measure the goodness of fit of the LinearMRF method. Figure~\ref{regressionChen} shows the regression line with coefficient of determination $R^2=0.3943$ on the dataset $Num=800$ when the RSE value is the smallest one of the LinearMRF in the ``Jackie Chan" community in Table~\ref{tab:RSE}. Figure~\ref{regressionZhou} shows the regression line with coefficient of determination $R^2=0.8115$ on the dataset $Num=200$ when the RSE value is the smallest one of the LinearMRF in the ``Zhi-Hua Zhou" community in Table~\ref{tab:RSE_zhou}. The results show the goodness of fit of the LinearMRF method.

Like~\cite{szabo2010predicting}, we use the Relative Squared Error (RSE) to evaluate the performance of the prediction models as shown in Equation~\ref{eq_RSE},
\begin{equation}\label{eq_RSE}
  RSE=\frac {\sum \limits_{k\in ST}(P(G,k)-real(G,k))^2} {\sum \limits_{k\in ST} (\overline{real(G,k)}-real(G,k))^2}
\end{equation}
where $P(G,k)$ is the predicted popularity of the topics of our methods and $real(G,k)$ is the real popularity of topics. $\overline{real(G,k)}=\frac{1}{|ST|} \sum \limits_{k\in ST}real(G,k)$ is an average of the real popularity of the topics.

\begin{table}[htbp]
  \centering
  \caption{Prediction Errors (Relative Squared Error) for the LinearMRF and EdgeMRF Models for Topics in the ``Jackie Chan" Datasets}
    \begin{tabular}{ccc}
    \hline
    \textbf{Dataset} & \textbf{LinearMRF}  & \textbf{EdgeMRF} \\
    \hline
    100   & 0.8564 & 0.7820 \\
    200   & 0.8595 & 0.7555 \\
    300   & 0.7710 & 0.7701 \\
    400   & 0.8732 & 0.7016 \\
    500   & 1.0536 & 0.7153 \\
    600   & 0.6935 & 0.6997 \\
    700   & 0.6845 & 0.7180 \\
    800   & 0.6437 & 0.7053 \\
    900   & 0.9148 & 0.7029 \\
    1000  & 0.6990 & 0.6764 \\
    \hline
    \end{tabular}%
  \label{tab:RSE}%
\end{table}
\begin{table}[htbp]
  \centering
  \caption{Prediction Errors (Relative Squared Error) for the LinearMRF and EdgeMRF Models for Topics in the ``Zhi-hua Zhou" Datasets}
    \begin{tabular}{ccc}
    \hline
    \textbf{Dataset} & \textbf{LinearMRF}  & \textbf{EdgeMRF} \\
    \hline
    1   & 0.8534 & 0.7371 \\
    10   & 0.8417 & 0.6957 \\
    50   & 0.7942 & 0.4453 \\
    100   & 0.6658 & 0.3748 \\
    150   & 0.5253 & 0.2112 \\
    200   & 0.5219 & 0.2607 \\
    \hline
    \end{tabular}%
  \label{tab:RSE_zhou}%
\end{table}

We evaluate the performance of the LinearMRF and EdgeMRF models in the ``Jackie Chan" dataset and ``Zhi-hua Zhou" dataset described in Section~\ref{sc_correlation}. In each test dataset, we randomly split the dataset to train dataset and test dataset with equal number of topics (the train dataset have one more topic when the number of topics in the dataset $Num$ is odd). Table~\ref{tab:RSE} and~\ref{tab:RSE_zhou} shows the prediction errors of the two models in the ``Jackie Chan" and ``Zhi-hua Zhou" datasets. We can find that only the RSE value 1.0536 of the LinearMRF model on dataset $Num=500$ is beyond 1.0 of the ``Jackie Chan" dataset. Other RSE values are all smaller than 1.0. In the ``Jackie Chan" dataset, the RSE values of LinearMRF in dataset $Num=600$, $Num=700$, $Num=800$ and $Num=1000$ are smaller than 0.7. The prediction errors of EdgeMRF are all less than 0.8 in the 10 datasets of ``Jackie Chan" community. In the ``Zhi-hua Zhou" dataset, the RSE value of EdgeMRF method even reaches 0.2112 when $Num=150$. The prediction errors of EdgeMRF are less than 0.5 when $Num$ is equal to or larger than 50. The results show that LinearMRF and EdgeMRF are effective in predicting popularity of topics in both the ``Jackie Chan" and ``Zhi-hua Zhou" dataset. We use Student's t-test to indicate statistically significant improvements of the EdgeMRF model over the LinearMRF model. The p-value of the EdgeMRF model over the LinearMRF model is 0.0691 in the ``Jackie Chan" dataset and 0.0014 in the ``Zhi-hua Zhou" dataset. It's a significant improvement in the significant level 0.1 and 0.01 in the two datasets. So the EdgeMRF model is statistically better than LinearMRF to predict popularity of topics.

\section{Discussion}\label{sc_discuss}

In this paper, we try to predict the popularity of topics which have not appeared in a microblogging website. Some existing works like TDT tasks have been done to detect and track topics. Some researches predict the popularity of online content based on their early popularity information. But these works predict the online content which has already happened. We are the first trying to predict the popularity of un-happened things.

The sentiment of a user on the key phrases of a topic can influence the decision of whether to discus about it. Based on this observation, we employ Markov Random Field (MRF) model and graph entropy model to compute the community sentiment energy of a topic, which is probably correlated with the real popularity of the topic. Experiments show the linear correlation between the community sentiment energy based on the MRF model and the real popularity of topics.

We propose two models LinearMRF and EdgeMRF to predict the popularity of topics. The relative squared errors (RSE) of EdgeMRF can reach about 0.7 in all the ten datasets of ``Jackie Chan" community. The value of RSE even reach 0.2112 in the dataset $Num=150$ of ``Zhi-hua Zhou" community. The results show the value of users' history sentiment on predicting the popularity of topics. It also shows the sentiment of a community can affect collective decision making of spreading a topic or not in the community. Our work can be used in many areas. For example, online social network marketing can use our work of predicting the popularity of topics to earn enough time in making plans for advertising. For potential candidate of government, it is very useful for them to find what the citizens care about for making helpful policies and earning the people's support.

The performance can be improved since our predicting models are simple and straightforward. We'll try to improve the models in future. The early population information and the structure of social network are also very helpful and we'll add this information to our model to improve the performance of predicting the popularity of topics in future.


\begin{acknowledgements}
This work is supported by National Natural Science Foundation of China (Grant No. 61502517) and the National Defense Science and Technology Project Funds (Grant No. 3101283).
\end{acknowledgements}

\bibliographystyle{spmpsci}      % mathematics and physical sciences
%\bibliography{HotTopic}   % name your BibTeX data base

%\begin{thebibliography}{}
% etc
%\end{thebibliography}

\end{document}